# Gas giant–like zonal jets in the laboratory

Daphné Lemasquerier ,[*] Benjamin Favier, and Michael Le Bars
*Aix Marseille Université, CNRS, Centrale Marseille, IRPHE, 13013 Marseille, France*

## I. INTRODUCTION

The origin, structure, and stability of gas giants' intense east-west winds (so-called zonal jets) are poorly understood [1]. In addition, another long-standing question is the origin of the dynamical difference between the gas giants' jets and those observed in terrestrial atmospheres and oceans [2]. From the fluid dynamics point of view, it is known that zonal jets emerge due to the variation of the Coriolis force with latitude, the so-called $\beta$ effect. However, we still lack a consensual model to account for the interaction between the large-scale slowly varying jets and the small-scale rapidly overturning turbulence. The turbulent nature of these flows and the wish to study their long-term behavior are two reasons motivating an experimental approach of this problem.

## II. EXPERIMENTAL SETUP

Our experiment consists in an improved version of the experiment described in Ref. [3], which allows for the spontaneous emergence of strong zonal jets in a forced and rotating turbulent flow. This setup is sketched in Fig. 1 and is described in detail in a separate paper [4]. A 1-m-diam and 1.6-m-high cylindrical tank filled with 600 l of water rotates at 1.25 rounds/s (75 rpm) to reproduce the effect of the planet's fast rotation. Due to the centrifugally induced pressure, the water upper free-surface adopts a paraboloidal shape leading to a varying height of fluid which generates a topographic $\beta$ effect:

$$|\beta| = \frac{2\Omega}{h}\frac{dh}{dr}, \qquad (1)$$

where $\Omega$ is the rotation rate, $h$ the fluid height, and $r$ the cylindrical radius. Physically, this $\beta$ effect acts as a source or sink of vertical vorticity following radial motions, due to the local conservation of angular momentum. To achieve a uniform $\beta$ effect, which is the simplest and the most thoroughly adopted configuration in theoretical studies, we use a sculpted bottom plate [Figs. 1 and 2(b)] so that the fluid height increases exponentially with radius. This compensation is accurate only at a fixed rotation rate of $\Omega = 75$ rpm $\approx 7.85$ rad/s and for a given fluid height at rest $h_0 = 0.58$ m. These parameters lead to an Ekman number of

$$\text{Ek} = \frac{\nu}{\Omega h_0^2} \approx 3.78 \times 10^{-7}, \quad (2)$$

where we assumed a constant kinematic viscosity for the water ($\nu = 10^{-6}$ m$^2$/s). Such a small Ekman number, associated with a free upper fluid surface, is crucial for experimentally generating zonal flows since dissipation by friction on the boundaries inhibits their formation.

---

[*]Corresponding author: lemasquerier.pro@protonmail.com

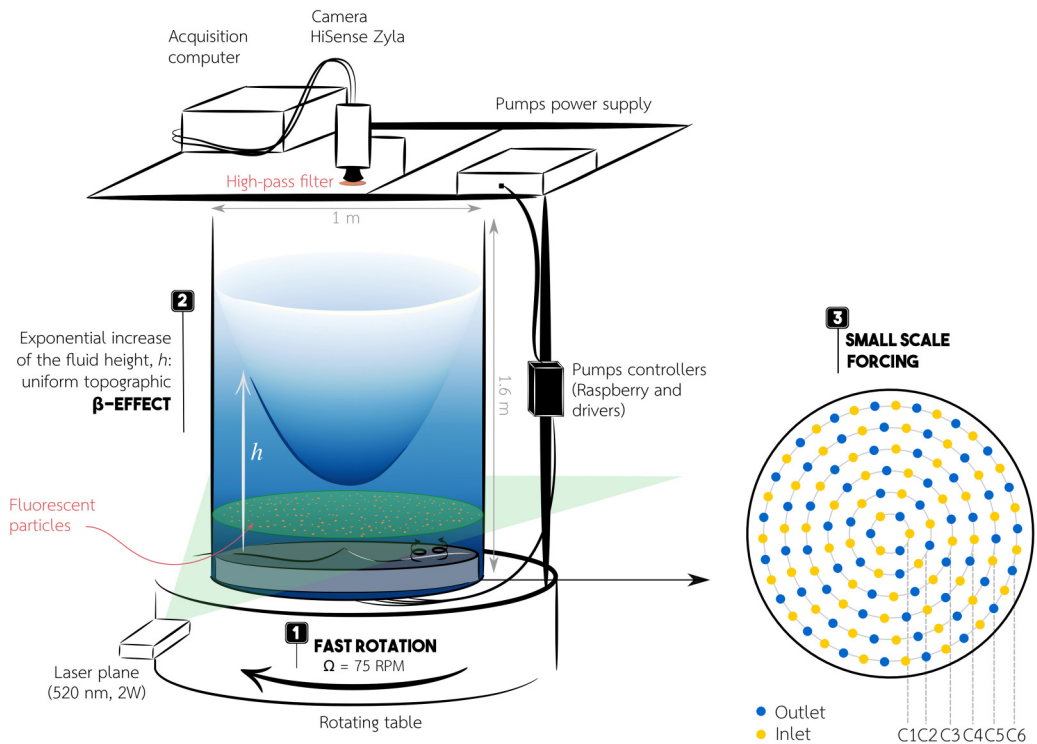

FIG. 1. Sketch of the experimental setup. The cylindrical tank is filled with ∼600 l of water and rotates at 75 rounds/min. A top lid (not represented) covers the tank so that the underlying air is also in solid-body rotation. All the devices are attached to the rotating frame. The polar lattice of the forcing is represented on the right.

Finally, we force small-scale fluid motions by using submersible pumps which circulate water through the curved bottom plate [Figs. 1 and 2(b)]. This forcing is a crude representation of the small storms, baroclinic instabilities, or convective motions occurring on the planets, which are all potential sources of energy for the zonal flow. The forcing is distributed on a polar lattice made of 128 holes, half of them being inlets (generating cyclones) and the other half outlets.

To measure velocity fields, time-resolving particle image velocimetry (PIV) measurements are performed in the rotating frame, on a horizontal laser plane located 9 cm below the center of the paraboloid [Figs. 1 and 2(a)]. The water is seeded with fluorescent red polyethylene particles of density 0.995 and 40–47 $\mu$m in diameter. Their motion is tracked using a top-view camera. The particles emit an orange light (607 nm), so using a high-pass filter on the lens allows us to filter out the green laser reflections on the free surface and tank sides [see Fig. 2(c)]. The velocity fields are deduced from these images using the MATLAB program DPIVSOFT [5]. Note that, due to the refraction of the laser plane by the tank sides, there are two shadow zones where measurements are not possible (see Fig. 3).

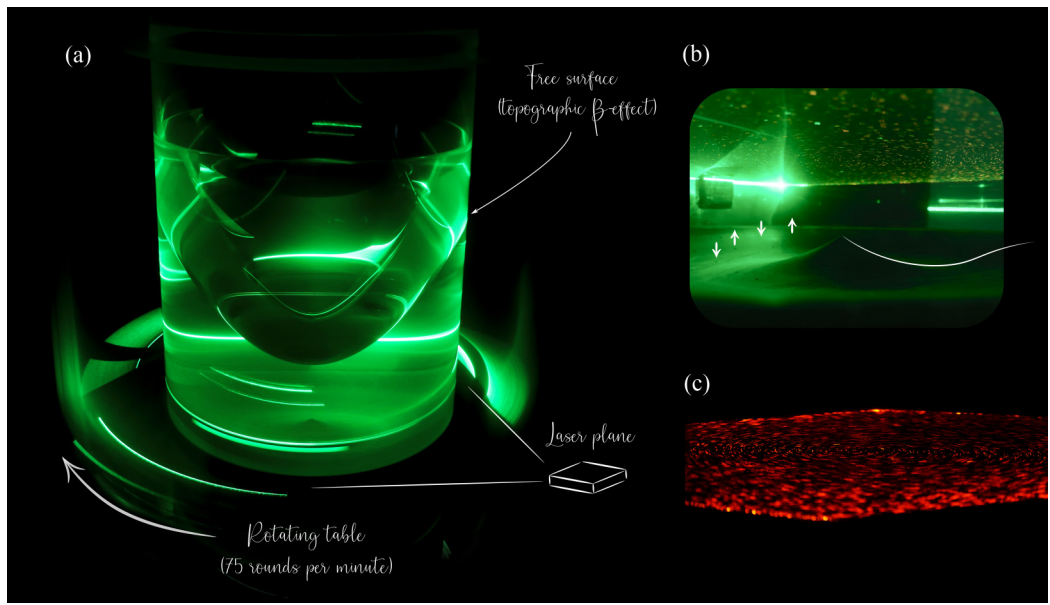

FIG. 2. (a) View of the experiment in solid-body rotation, with the laser turned on. The paraboloidal free surface is visible, as well as the central peak of the bottom plate through which the forcing is performed. The laser plane is located in between the center of the paraboloid and the bottom plate. (b) View below the laser plane, showing part of the bottom plate. It is curved and has the shape of a curly bracket. The tank's water circulates through the holes drilled into this plate due to submersible pumps located beneath it. (c) View of the laser plane with an orange filter. Only the fluorescent particles are visible, which eliminates unwanted reflections. The original poster is available at https://doi.org/10.1103/APS.DFD.2019.GFM.P0015.

An experimental run is as follows. We gradually increase the table rotation rate up to 75 rpm. Once solid-body rotation is reached, which takes approximately 45 min, i.e., $\sim 13\tau_E$, where $\tau_E = \Omega^{-1}\text{Ek}^{-1/2}$ is the Ekman spin-up timescale, the fluid free surface is still and paraboloidal, with a height of 20 cm at the center and 96 cm at the border of the tank. We then turn on the six forcing pumps simultaneously, in a stationary state, and let the flow self-organize. For a typical run, we record images for 60 min, corresponding to $4500 t_R$, where $t_R = 2\pi/\Omega = 0.8$ s is the rotation period.

## III. TOP-VIEW OBSERVATIONS

With those three ingredients, the fast rotation, the $\beta$ effect, and the small-scale forcing, we indeed observe the spontaneous generation of multiple zonal jets, going successively in prograde and retrograde directions (Fig. 3). Additionally, these jets are instantaneous, meaning that we do not need to perform any time averaging to reveal them.

Depending on the forcing amplitude, we obtain two different regimes of zonal jets. At low amplitude, we observe the generation of five prograde and five retrograde jets (regime I). Figure 3(a) shows streamlines of the flow obtained after a statistically steady state is reached, which takes only $\sim 100$ rotation times in this regime. The jets coexist with numerous small vortices associated with the forcing. At high forcing amplitude, regime I initially develops before merging and evolving towards broader jets of higher amplitude (regime II), as illustrated in Fig. 3(c). For the chosen experiment, three prograde and three retrograde jets were obtained, but their number and position may vary for the same forcing. In regime II, the small-scale vortices merge into larger ones, between which the jets still meander. This spontaneous transition from regime I to regime II is slow, and the statistically steady state is obtained after a transient of about $800 t_R$.

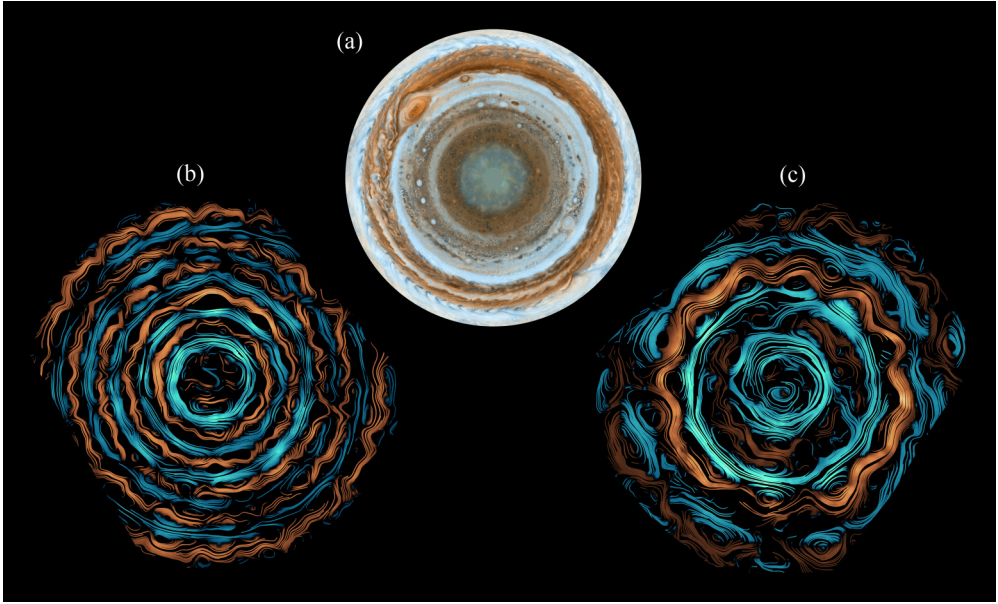

FIG. 3. (a) Jupiter view from the south pole. Credits: NASA/JPL/Space Science Institute. A top view of our experiment is equivalent to a view of Jupiter from the pole, except that in our experiment the $\beta$ effect is uniform. (b) Regime I. The pumps' power, from the inner to the outer ring, is $P_i = \{14, 20, 46, 72, 100, 100\}\%$ of their maximum power. We represent streamlines computed from the experimental velocity fields measured by PIV at a time $t = 840 t_R$. The velocity field used to generate this image is averaged over 5 s. The colors represent the sign of the azimuthal velocity, with positive (prograde) in red and negative (retrograde) in blue. The color scale ranges from $-0.8$ to $0.8$ cm/s. (c) Regime II. The forcing is $P_i = \{26, 33, 60, 80, 100, 100\}\%$. The same data as in (b) are plotted, except that the velocity field is computed at time $t = 1780 t_R$ and the color scale ranges from $-4.0$ to $4.0$ cm/s. There is thus a factor 5 between the two color scales: The jets are of significantly higher amplitude in regime II. The original poster is available at https://doi.org/10.1103/APS.DFD.2019.GFM.P0015.

Both regimes correspond to low-Rossby-number quasigeostrophic dynamics, the local Rossby number (vorticity over twice the rotation rate) being inferior to 0.1 in both cases. The physical origin of this transition and the associated bistability—the two regimes can coexist for the same forcing amplitude—is due to the resonance of the directly forced Rossby waves which are advected by the zonal flow. This process is investigated in detail in a separate paper [4]. We argue that regime II, where the jets grow in size and amplitude and choose their position almost independently of the forcing pattern, is the one relevant for applications to the gas giants.

## ACKNOWLEDGMENTS

The authors acknowledge funding by the European Research Council under the European Union's Horizon 2020 research and innovation program through Grant No. 681835-FLUDYCO-ERC-2015-CoG. The authors are most grateful to E. Bertrand and W. Le Coz for their help and ingenuity during the conception and building of the experiment. We also thank J.-J. Lasserre for his help in setting up and calibrating the PIV system.